%% file: main.tex
\documentclass[12pt, letterpaper]{article}
\usepackage[margin=2cm]{geometry} 
\usepackage{lmodern}
\usepackage[scr=rsfs]{mathalpha}
\usepackage{cite}
\usepackage{amssymb}
\usepackage{amsmath}
\usepackage{graphicx}
\usepackage{physics}

\newcommand{\ud}{\dd}

\newcommand{\ue}{\mathrm{e}}

\usepackage{nicefrac}
\usepackage{url}

\title{Statistical Mechanics of Unconfined Systems: Challenges and Lessons}

\author{Bruno Arderucio Costa$^1$, Pedro Pessoa$^2$ \\
\\
$^1$Institute for Theoretical Physics - UNESP \\ S\~ao Paulo, SP - Brazil\\
$^2$Department of Physics, University at Albany - SUNY \\  Albany, NY - USA}

\date{\today}

\begin{document}
\maketitle
\abstract{Motivated by applications of statistical mechanics in which the system of interest is spatially unconfined, we present an exact solution to the maximum entropy problem for assigning a stationary probability distribution on the phase space of an unconfined ideal gas in an anti-de Sitter background. Notwithstanding the gas's freedom to move in an infinite volume, we establish necessary conditions for the stationary probability distribution to be normalizable. As a part of our analysis, we develop a novel method for identifying dynamical constraints based on local measurements. With no appeal to \emph{a priori} information about globally-defined conserved quantities, it is thereby applicable to a much wider range of problems.}


\input{text}

\section*{Acknowledgements} The authors wish to thank Ariel Caticha for insightful conversations. BAC received funding from Conselho Nacional de Desenvolvimento Cient\'ifico e Tecnol\'ogico under the process 162288/2020-4.

\bibliography{Referencias}


\end{document}

%% file: text.tex
\section{Introduction}
Despite its overarching success, the statistical mechanics of Gibbs~\cite{Gibbs02} falls short in its description of spatially unconfined systems. In the canonical ensemble, a classical gas of $N$ particles is described by a Gibbs probability distribution $f(y),\ y \in \mathcal Y$ defined over the phase space $\mathcal Y = \{( q_i, p_i)\}$, for $i \in \{1, 2, \ldots, N\}$ where $q_i$ and $p_i$ are the position and momentum for the $i$-th particle). The Gibbs distribution is of the form $f(y) \propto \ue^{-\beta H(y)}$, where $\beta$ is the unit-corrected inverse temperature (throughout this paper, we choose units in which $k_B=c=1$) and $H$ is the system Hamiltonian. In cases where the gas is unconfined, meaning the set of allowed positions is unlimited, the canonical ensemble may not lead to a valid probability distribution. The normalization condition in this ensemble is

\begin{equation}
    \int\ud y\ \ue^{-\beta H(y)}<\infty.
    \label{partitionF}
\end{equation}

One usually can write the Hamiltonian in the block-diagonal form

\[H(y)=\sum_{ij} F(q_{ij})+\sum_i \frac{p_i^2}{2m},\]
where $q_{ij} = q_i-q_j$. Condition~(\ref{partitionF}) is manifestly not satisfied whenever $F$ falls off to infinity slower or as slow as $1/|q_{ij}|$. In particular, the normalization condition is not met for Newton/Coulomb-type interactions or a constant in $q_{ij}$ (i.e., the ideal gas).

In the kinetic theory, the situation is no better. As already pointed out in ref.~\cite{Lifshitz79}, solutions to the equilibrium Boltzmann equation are spatially homogeneous in the absence of external forces.

There are, however, examples of unconfined systems of physical interest, for example, an unconfined gas of particles interacting through gravity, such as a star or a galaxy cluster. The statistical mechanics of such a system has been studied before and, to the best of our knowledge, every study thus far on the self-gravitating gas (or a gas of Coulomb-interacting particles \cite{Davis20}) requires confinement demanding that the particles are only allowed to occupy a limited set of position coordinates, i.e., confined in a box~\cite{Ahmad02,Ahmad06,deVega02,deVega02b}. 
This approach allows one to bypass the normalization problem by restraining the integral~(\ref{partitionF}) over the coordinates $q_i$ to a compact domain of volume $V$.

In spite of its effectiveness for astrophysical applications, progress in this direction leaves fundamental questions about the awkwardness of dealing with unconfined systems unanswered. For example, whether or not it is possible to display a system that reaches thermodynamical equilibrium while kinematically allowed to occupy an infinite volume. This article answers this question affirmatively.

\section{Background}

The work of Jaynes~\cite{Jaynes57} solidified the statistical mechanics of Gibbs through what is currently known as the method of maximum entropy (MaxEnt). Further developments --- starting with Shore and Johnson \cite{Shore80} --- present MaxEnt as a method to select and update probability distributions $f$ when information about the system is revealed by maximizing the functional

\begin{equation}\label{entropy}
S[f] = - \int \dd y \ f(y) \log \left( \frac{f(y)}{\varphi(y)} \right) \ ,
\end{equation}
where $\varphi$ is a prior distribution,  with constraints meant to represent the information at hand. Usually these constraints are in the form of expected values for a series of functions $a_\mu(y)$, namely sufficient statistics. The $f$ that maximizes \eqref{entropy} under normalization, $\int \dd y \ f(y)=1$, and a set of $n$ expected value constraints, $A_\alpha = \int \dd y \ f(y) a_\alpha(y)$ for $\alpha \in \{1,\ldots,n\}$, is the Gibbs distribution

\begin{equation}\label{Gibbs}
f(y|\beta) = \frac{\varphi(y) }{Z} \exp{- \sum_\alpha \beta_\alpha a_\alpha(y)} \ ,
\end{equation}
where $Z(\beta)$ is a normalization factor, $Z = \int \dd y \ q(y) \exp{- \sum_\alpha \beta_\alpha a_\alpha(y)}$, and $\beta_\alpha$ are the Lagrange multipliers associated with the expected value constraint $A_\alpha$.
Fundamental factors in statistics imply that only entropies of the form \eqref{entropy} are appropriate for updating information --- see e.g. \cite{Shore80,Caticha,Vanslette17}. It follows from \eqref{Gibbs} that the Lagrange multipliers and the expected values are related through $A_\alpha =-  \pdv{}{\beta_\alpha} \log Z$. 

The canonical ensemble of statistical mechanics assumes a system of interest connected to a heat bath, namely a much larger system that exchange energy with the system of interest (see e.g.~\cite{Caticha}). In this model, the only sufficient statistic is given by the Hamiltonian of the system $a_1(y) = H(y)$.  
In the description given by the canonical ensemble, one may say that, when the variational problem of extremizing \eqref{entropy} with the expected value of energy constraint cannot be solved, it means the system does not reach thermal equilibrium. This is the case for the self-gravitating gas, as explained in the introduction. It then follows that, in order to properly describe a thermodynamical system held to itself by gravity, we should evolve MaxEnt into the dynamical description, allowing the system to evolve in time.

Describing time evolution, in our context, means studying how the function $f$ changes when a particle moves along its worldline (i.e., its trajectory in spacetime). Since we are dealing with massive particles, we can always use the proper time $\tau$ to parametrize their worldlines. Note that a change in the position $x^a$ is accompanied by a change in the tangent vectors to the trajectory, which are proportional to the momentum $p^a$. To keep our notation as simple as possible, we implicitly interpret $f$ below as a function of $\tau$ using the composition $f(x^a(\tau),p^a(\tau))$.

In statistical mechanics, a very common choice for the prior $\varphi(y)$ in eq.~\eqref{entropy} is a uniform distribution over $\mathcal Y$. This choice implements a symmetrical ignorance about the system: in the lack of information favouring a point over any other, it is fair to demand the prior to reflect the homogeneity. Under the understanding that time evolution preserves information, we have, adopting a uniform prior,

\begin{equation}
   \dv{}{\tau}S[f(x^a,p^a)]=\fdv{S}{f}\times\dv{f(x^a,p^a)}{\tau}. 
\end{equation}
The first factor is, by eq.~\eqref{entropy}, nonzero. Hence, $\dv{S}{\tau}=0$, implying $\dv{f}{\tau}=0$ and consequently~\cite{Arnold74},

\begin{equation}
    \left(\frac{1}{m}p^\mu\pdv{}{x^\mu}+\dv{p^\mu}{\tau}\frac{\partial}{\partial p^\mu}\right)f=0,
    \label{Liouville}
\end{equation}
where $\tau$ is the proper time along the integral curves of $p^a$ and $m$ is the pass of a particle of the gas.

Equation~(\ref{Liouville}) provides us with a necessary condition the function $f$ has to fulfil in order to extremize the entropy.

Our case of study is the anti-de Sitter spacetime, a maximally symmetric space with negative curvature. As for any space of constant curvature, its curvature tensor is fully determined by its scalar curvature $R$. In our case, $R=-12/\Lambda^2$, where $\Lambda$ is the \emph{anti-de Sitter length}. Consequently, the limit $\Lambda\to\infty$ recovers the local geometry of Minkowski spacetime. The anti-de Sitter space can be thought of as a solution for the vacuum Einstein's equation with a cosmological constant $-3/\Lambda^2$.
For a  lengthy discussion about this spacetime, we suggest the refs.~\cite{HawkingEllis, MisnerThroneWheeler}. For the purposes of the discussion presented here, note that although it is not globally hyperbolic, the anti-de Sitter spacetime can be foliated by a family of spacelike surfaces of infinite volume.

\section{Solutions of the Boltzmann equation in anti-de Sitter spacetime}
We work with the universal covering space of an anti-de Sitter spacetime described by the metric

\begin{equation}
    \ud s^2=-\left(1+\frac{r^2}{\Lambda^2}\right)\ud t^2+\left(1+\frac{r^2}{\Lambda^2}\right)^{-1}\ud r^2+r^2\ud\theta^2+r^2\sin^2\theta\ud\phi^2 \ ,
    \label{ads}
\end{equation}
with $t\in\mathbb R$, $r\in [0,\infty)$, $\theta\in [0,\pi]$ and $\phi\in [0,2\pi)$. In these coordinates, it is clear that $\xi^a=\left(\frac{\partial}{\partial t}\right)^a$ is a timelike Killing vector field.

The collisionless Boltzmann equation for an ideal gas in a generic curved spacetime follows from applying the geodesic equation to eq.~(\ref{Liouville}) and reads~\cite{Ellis83}

\begin{equation}
    \left(p^\mu\frac{\partial}{\partial x^\mu}-\Gamma^\mu_{\nu\rho}p^\nu p^\rho\frac{\partial}{\partial p^\mu}\right)f(x^\lambda,p^\lambda)=0 \ ,
    \label{boltzmanneq}
\end{equation}
where $\Gamma^\mu_{\nu\rho}$  are the Christoffel symbols.
Eq.~(\ref{boltzmanneq}) assumes the motion to be geodesic. This means that the ideal gas is an inbuilt assumption for its validity.

We shall seek spherically symmetric solutions of eq.~(\ref{boltzmanneq}) on the metric~(\ref{ads}). For these solutions $f(x^\lambda,p^\lambda)$ can be written as a four-variable function  $\tilde f(t,r,p^0,p^r)$, which, when substituted in \eqref{boltzmanneq}, yields

\begin{equation}
p^0\frac{\partial\tilde f}{\partial t}+p^r\frac{\partial\tilde f}{\partial r}-\frac{2r}{r^2+\Lambda^2}p^0p^r\frac{\partial\tilde f}{\partial p^0}-\left[\frac{r(r^2+\Lambda^2)}{\Lambda^4}(p^0)^2-\frac{r}{r^2+\Lambda^2}(p^r)^2\right]\frac{\partial\tilde f}{\partial p^r}=0 \ ,
    \label{fullBEAdS}
\end{equation}
restricted to the submanifold with $p^\theta=p^\phi=0$.

Equation~(\ref{fullBEAdS}) admits a separation of variables as
\begin{equation}
    \tilde f(t,r,p^0,p^r)=T(t)\mathcal S(r,p^0,p^r),
    \label{defts}
\end{equation}
where $T$ and $\mathcal S$ are functions to be determined.

The equation for $t$ has a simple exponential solution $T(t) \propto \ue^{-t/t_c}$, while the equation for $\mathcal S$ cannot be solved using elementary methods unless the separation constant $1/t_c=0$, in which case the solutions are referred to as \emph{stationary}. It is important to bear in mind that the condition $\dd f/\dd\tau=0$, which follows directly from the maximization of the entropy, does not necessarily entail $\dd f/\dd t=0$.

By direct substitution into eq.~\eqref{defts}, we can see that

\begin{equation}
    \mathcal S=h\left(p^0\left(1+\frac{r^2}{\Lambda^2}\right),-\left(1+\frac{r^2}{\Lambda^2}\right)(p^0)^2+\left(1+\frac{r^2}{\Lambda^2}\right)^{-1}(p^r)^2\right) \ ,
    \label{generalsoln}
\end{equation}
for any function $h\in\mathscr C^2(\mathbb R^2)$ solves eq.~(\ref{fullBEAdS}).

The arguments of the function $h$ have a simple physical interpretation. The second is $g_{\mu\nu}p^\mu p^\nu$ when $p^\theta=p^\phi=0$. This quantity is merely a constant, namely $-m^2$, where $m>0$ is the mass of the particles constituting the gas. The first argument is an integral of the motion,

\begin{equation}
\varepsilon\equiv-p^a\xi_a,   
\label{epsilon}
\end{equation}
the conserved energy of a particle~\cite{Wald84}.

This result is not surprising. Indeed, progress was made in the 60s for simultaneously solving Einstein's equations and the equations of motion for matter consisting of a collection of particles. When a Hamiltonian can be defined for such systems, Liouville's equation is a consequence of Einstein's field equations. Fackerell~\cite{Fackerell68} and Ehlers \emph{et al.}~\cite{Ehlers68}, based on an analysis of the characteristic equations of (\ref{boltzmanneq}), identified that the general solutions of eq.~\eqref{Liouville} can be written as functions of the conserved quantities. As a matter of fact, it follows that any function of first integrals of motion of the individual particles automatically satisfies the stationary Liouville equation~\eqref{Liouville}. This shows that not only is eq.~\eqref{generalsoln} the general solution for the stationary spherically symmetric case, but also a \emph{particular solution for the general stationary case}, i.e. when $f$ is allowed to depend on the angular variables as well as having non-zero arguments $p^\theta$ and $p^\phi$.

Note that, unlike refs~\cite{Fackerell68,  Ehlers68}, we are not interested in finding solutions to the full gravitational problem. Rather, we wish to preserve the analogy with the flat spacetime version of the problem, thereby neglecting any gravitational effects caused by the gas or its parts.

The nature of the constant $-m^2$ is different to the nature of $\varepsilon$. The latter is a first integral of the motion, i.e., a function of phase space that is a constant along the geodesics. The former is more conveniently thought of as a kinematic invariant. 
For simplicity, we henceforth focus on solutions with vanishing angular part of the four-momentum but we shall return to this point in due time.
Hereafter, we use the $-m^2$ invariant to restrict $\mathcal S$ to the co-dimension one subspace $\mathcal P$ of the section of $\mathcal Y$  and write $\mathcal S=\tilde h(\varepsilon)$ for a single-variable function $\tilde h$ of $\varepsilon$, defined in~\eqref{epsilon}.

Naturally, not all solutions of the form $\mathcal S=\tilde h(\varepsilon)$ are normalizable. The ubiquitous normalization constraint reads~\footnote{Naturally, additional constraints may be present in the MaxEnt problem. The integral only ranges for non-negative values of $p^0$ because the condition that $p^a$ is future-directed entails $p^0>0$.}

\begin{equation}
\int_m^\infty\ud  p^0\int_0^\infty\ud r\ \frac{r^2}{\sqrt{1+r^2/\Lambda^2}}\ \tilde h(\varepsilon(p^0,r))<\infty \ ,
\end{equation}
leads to an asymptotic (large $r$) behaviour of $\tilde h$ falling off at least as quickly as 

\begin{equation}
    \tilde h(\varepsilon)\sim\varepsilon^\alpha\quad \text{with}\quad \alpha<-1.
    \label{falloff}
\end{equation}

Had we allowed for non-zero angular components of the four-momentum, our conclusions would not be challenged. Owing to the parity-reversal symmetry of the metric, we study the geodesics on the equatorial plane $\theta=\nicefrac{\pi}{2}$ without loss of generality. Instead of eq.~\eqref{generalsoln}, we now have $\mathcal S=h(\varepsilon,r^2 p^\phi,m^2)$ since $\ell\equiv p^a\psi_a=r^2p^\phi$ is another independent integral of the motion associated with the Killing field $\psi^a=\left(\nicefrac{\partial}{\partial\phi}\right)^a$. The constant of the motion $\ell$ can be thought of as the angular momentum of a particle per unit mass. Using an entirely analogous reasoning, we construct the two-variable function $\tilde h(\varepsilon,\ell)$, which behave as $\tilde h\sim\varepsilon^\alpha\ell^\gamma$ for large values of the radial coordinate. The normalization conditions for $p^0$, $p^\phi$, and $r$ demand, respectively, that $\alpha<-1$, $\gamma<-1$, and $\alpha+\gamma<-2$, the last of which, of course, being redundant. This shows that there are normalizable solutions to the Liouville equation also when the angular part of the four-momentum is free to assume any kinematically-allowed value. In other words, normalized distributions in the momenta are also automatically normalizable with respect to the configuration space variables.

\section{Constraints and MaxEnt}
Because the format of $f$ as $\tilde h(\varepsilon,\ell)$ was derived from imposing $\dd S/\dd\tau=0$, we have thus far deduced a necessary condition for $f$ to extremize the entropy, but by no means sufficient. We now show that among solutions obeying eq.~(\ref{falloff}), there are extremes of the entropy subjected to physically relevant constraints.

We adapt the arguments from Caticha~\cite{Caticha} to model a ``measuring device'' one can use to gain information from the gas and subsequently utilize the acquired information to set constraints to the variational problem. The procedure we describe below uses local measurements to impose constraints to the system of interest. This idea can be used in other problems in general relativity where the use of global conditions are not as straightforward or uniquely defined as they are in non-relativistic mechanics.

Our measuring device $D$ consists of another ideal gas; this time confined into a box whose walls are perfectly permeable to the unconfined gas' particles, but impenetrable to the particles of $D$. As with any measuring device, $D$ has to interact with the system of interest. We model this interaction by elastic collisions between particles of $D$ and particles of the unconfined gas~\footnote{An idea similar to ours was discussed in flat spacetime by Kubli and Herrmann in ref.~\cite{Kubli20}, where the authors elegantly described the interaction between their measuring device and the system of interest through an extra interaction term in the Hamiltonian. Our approach instead appeals to kinetic theory, is more general, and agrees with their results for relativistic particles in flat spacetime.}. $D$ itself being an ideal gas, its constituents follow geodesics between successive collisions and hence cannot absorb energy from the gravitational field. In addition, we make the box large enough so that the thermodynamic limit can be applied to the gas $D$, but small in comparison to $\Lambda^{3}$ so that $D$ possesses a well-defined total energy. If this energy changes by an amount $\Delta E$, a small subset $I\subset D$ of the particles of the measured gas must have suffered a change of $\sum_{i\in I}\varepsilon_i=-\Delta E$. Therefore,

\begin{equation}
    \mathcal E\equiv\sum_{i\in D}\varepsilon_i+E
\end{equation}
is an \emph{unknown} constant. It is important to emphasize that $\mathcal E$ does not have the same status as the ``total energy'' in the non-relativistic theory. Its physical interpretation is neither as straightforward nor easily relatable to the Arnowitt-Deser-Misner conserved mass of the spacetime, which would include contributions of pure gravity, as well as the backreaction of the gas onto the metric.

With the knowledge of the value of $E$ given by the measurement in $D$~\footnote{This measurement can follow any procedure one would use to measure the total energy of an ideal gas confined in a box sitting on a desk. For example, one can put it on a set of weighing scales.}, we impose the normalization constraint and

\begin{equation}
\int_{\mathcal P}\left\{\mathcal E-\varepsilon(y)\right\}f(y)\ud y=E,
    \label{enconst}
\end{equation}
which leads to the Gibbs distribution

\begin{equation}
f(y)=\frac{1}{Z}\ue^{-\beta\varepsilon(y)}\propto\ue^{-\beta\ p^0\  (1+\nicefrac{r^2}{\Lambda^2})}.
    \label{GibbsAdS}
\end{equation}
Here, for sake conformity with the notation from section 2, we changed the notation from discrete sums $\sum_i$ to integrals $\int_{\mathcal P}$.

Clearly eq.~(\ref{GibbsAdS}) respects condition~(\ref{falloff}) and consequently the normalization constant $Z$ is finite. A substitution of (\ref{GibbsAdS}) into (\ref{enconst}) reveals

\begin{equation}
E=\frac{\partial}{\partial\beta}\log Z+\mathcal E.
\label{curlyE}
\end{equation}

The partition function, as for any ideal gas, factors out~\footnote{If one wishes to impose that the particles are indistinguishable from one another, the extra factor $1/N!$ is needed to account for the actual phase space being the quotient of all possible permutations of the pairs $(q_i,p_i)$ in the product space of the one-particle phase spaces. The introduction of this factor is immaterial for our present discussion.} as $Z=\zeta^N$ and can be calculated explicitly:

\begin{multline}
    \zeta =4\pi\int_m^\infty\ud p^0\int_0^\infty\ud r\ \frac{r^2\exp\left[-\beta p^0\left(1+\frac{r^2}{\Lambda^2}\right)\right]}{\sqrt{1+\frac{r^2}{\Lambda^2}}}=\\
    =\frac{2\pi\Lambda^3}{\beta}\ue^{-\beta m/2}\left[(1+\beta m)K_0\left(\frac{\beta m}{2}\right)-\beta mK_1\left(\frac{\beta m}{2}\right)\right],
    \label{zeta}
\end{multline}
where $K_\nu(z)$ is the modified Bessel function of the second kind. The flat spacetime is recovered from eq.~(\ref{zeta}) in the limit $\Lambda\to\infty$,  in which case the partition function diverges as it should.

Remarkably, substituting eq.~\eqref{zeta} into~\eqref{curlyE}, we see that $\mathcal E$ does not depend on the anti-de Sitter length $\Lambda$, suggesting that the equations of state so obtained could accurately describe the gas in flat space as well. A calculation of the entropy from eqs.~\eqref{GibbsAdS} and~\eqref{zeta} yields a term independent of $\Lambda$ and another depending on $\Lambda$ logarithmically. The latter, in the case of $N$ indistinguishable particles, can be made small in comparison with the former for sufficiently large $N$.

\section{Discussion and prospects}
With the example of an ideal gas in anti-de Sitter spacetime, we hope to have elucidated that it is not the kinematical possibility to occupy an infinite volume that prevents a system from having a well-posed equilibrium probability distribution. We emphasize that the results of section 3 are independent of the choice of constraints made in section 4. Specifically, all normalizable stationary solutions of the MaxEnt problem supplied with whatever set of constraints must obey condition~\eqref{falloff}.

We interpret the divergence of the partition function of the ideal gas in flat spacetime as follows. When the phase space $\mathcal Y$ describes a collection of a fixed number of particles and the partition function factors out as $Z=\zeta^N$ as above (see also footnote 4), if particles of arbitrarily low energies can visit any point of the $q_i$ section of $\mathcal Y$, the spatial integral of the partition function is forced to sample the infinite volume in all its glory.

The convergence in anti-de Sitter space is explained after describing its radial timelike geodesics. They can be calculated directly from \eqref{ads} requiring $g_{ab}p^ap^b=-m^2$ together with the condition that $\varepsilon$ is a constant along them. The results are sinusoidal functions $r$ of the proper time $\tau$~\footnote{For reference, eliminating $\ud t/\ud\tau$ in favour of $\varepsilon$ and substituting on the dispersion relation, we obtain $(\ud r/\ud\tau)^2=\varepsilon^2-1-r^2/\Lambda^2$. Taking a derivative of this expression with respect to $\tau$ we get the equation of motion of a simple harmonic oscillator, $\ud^2r/\ud\tau^2+\Lambda^{-2}\ r=0$. We note that the frequency of oscillation is independent of the particle's energy.}. A more geometrical interpretation is that timelike geodesics are oscillatory: they continue indefinitely departing from a point and reconverging to another, never reaching infinity. Higher amplitudes are accompanied by higher values of $\varepsilon$ so that the $\zeta$-integral only samples the infinite spatial volume ``weighed down'' by an appropriate decreasing function. This view is endorsed by the realization that eq.~\eqref{zeta} is divergent on the limit $m\to0$: null geodesics do not share the oscillatory behaviour of their timelike counterparts. Rather, they reach the null infinity $\mathscr{I}$ regardless of their energy $\varepsilon$.

In this respect, one can think of the anti-de Sitter geometry acting on massive particles as an ``external field'' binding the system together. Let us digress and illustrate this analogy with an example: an unconfined, non-relativistic ideal gas in Minkowski spacetime interacting with a spherically-symmetric harmonic well is described by the Hamiltonian $H=\sum_i \nicefrac{p_i^2}{2m}+\nicefrac{1}{2}\ m\omega^2 r_i^2$, where $r_i$ is the radial distance from the $i$-particle and the centre of the well. For this system, the sum of the energies of each individual particle \emph{is} the total energy and the canonical ensemble exists. In this ensemble, the probability distribution has the same Gaussian dependence on the radial variable as in eq.~\eqref{GibbsAdS}. A direct integration gives

\begin{equation}
    \zeta=\frac{8\pi^3}{\beta^3\omega^3}.
\end{equation}
In this simple system, the frequency $\omega$ acts like the reciprocal of the anti-de Sitter length $\Lambda$. The expected value of the total energy per particle, $-\pdv{}{\beta}\log\zeta=\nicefrac{3}{\beta}$, does not depend on $\omega$, in direct analogy with our earlier observation that for the gas in anti-de Sitter, $\mathcal E$ is independent of $\Lambda$. 

The static patch of the de Sitter spacetime permits another straightforward test case. This patch is described by the metric~(\ref{ads}) under the replacement $\Lambda^2\to-\Lambda^2$. The separation of variables for the dynamical case discussed in the paragraph below eq.~\eqref{fullBEAdS} applies equally well for the de Sitter case. However, the static patch, being part of the steady-state universe, drives a family of geodesic-following particles apart from one another. Therefore, based on the interpretation above, we do not anticipate the existence of stationary normalizable solutions whatsoever. This is indeed the case. The convergence of the spatial integral of $\tilde h(\varepsilon)$ as $r$ approaches $\Lambda$ requires $\tilde h\sim\varepsilon^\alpha$ with $\alpha\geq\nicefrac{1}{2}$, whilst the convergence condition on the integral over $p^0$ requires $\alpha<-1$. The impossibility of satisfying both these conditions at once proves that no function $\tilde h$ can be normalized. Again, this conclusion would not be altered if we allow the angular momentum $\ell$ to be free. This can be seen immediately from the realization that an extra factor of $\ell^\gamma$ is regular on $r=\Lambda$ for all $\gamma$, thereby playing no role when studying the possible divergences near the de Sitter horizon at $r=\Lambda$.

In summary, this paper undertook two missions. First, it established general necessary conditions for the existence of normalizable stationary solutions for the maximum entropy problem for a relativistic ideal gas in anti-de Sitter spacetime. Such conditions show that normalizable solutions may exist even when there is no ``wall'' or external forces confining a gas in place. Second, it proposed a type of constraints based on local measurements that can have applications transcending the study of unconfined systems.

The success of our programme for ideal gases suggests it can also be successful for long-range interacting gases. It would be instructive to compare the dependence of thermodynamic properties of, say the self-gravitating star, on the ``regularization procedure''. That is to say, if the equations of state of a self-gravitating star in anti-de Sitter spacetime with very large $\Lambda$ agree with the large-$V$ results from refs.~\cite{deVega02,deVega02b}~\footnote{We do not expect the results thereby obtained to agree with the ones from refs.~\cite{Ahmad02,Ahmad06} because these references introduce an interaction cut-off, which ultimately leads to extensivity, a property not expected to hold when the interaction is Newtonian until infinity~\cite{PP20}.}. Even more importantly, special attention must be given to any thermodynamic potential that turns out to be independent of $\Lambda$.